\documentstyle[titlepage,12pt]{article}
\def\openone{\leavevmode\hbox{\small1\kern-3.3pt\normalsize1}}

\newcommand{\pr}[2]{{\em Phys.\ Rev.\ }{\bf #1}{(#2)}}
\newcommand{\prd}[2]{{\em Phys.\ Rev.\  D\ }{\bf #1}{(#2)}}
\newcommand{\prl}[2]{{\em Phys.\ Rev.\ Lett.\ }{\bf #1}{(#2)}}
\newcommand{\pl}[2]{{\em Phys.\ Lett.\ }{\bf #1}{(#2)}}
\newcommand{\pla}[2]{{\em Phys.\ Lett.\ A\ }{\bf #1}{(#2)}}

\newcommand{\epl}[2]{{\em Europhys.\ Lett.\ }{\bf #1}{(#2)}}
\newcommand{\jmp}[2]{{\em J.\ Math.\ Phys.\ }{\bf #1}{(#2)}}

\newcommand{\prsl}[2]{{\em Proc.\  Roy.\  Soc.\ (London)\  A\ }{\bf #1}{(#2)}}

\newcommand{\mpl}[2]{{\em Mod.\ Phys.\ Lett.\ }{\bf #1}{(#2)}}
\newcommand{\ijmp}[2]{{\em Int.\ J.\ Mod.\ Phys.\ }{\bf #1}{(#2)}}
\newcommand{\rmp}[2]{{\em Rev.\ Mod.\ Phys.\ }{\bf #1}{(#2)}}
\newcommand{\apny}[2]{{\em Ann.\ Phys.\ (N.Y.)\ }{\bf #1}{(#2)}}

\newcommand{\jp}[2]{{\em J. \ Phys.\  }{\bf #1}{(#2)}}

\newcommand{\nc}[2]{{\em Nuovo\ Cim. \ }{\bf #1}{(#2)}}

\newcommand{\lnc}[2]{{\em Lett.\  Nuovo \ Cim.\ } {\bf #1}{(#2)}}
\newcommand{\fp}[2]{{\em Found.\  Phys.\ } {\bf #1}{(#2)}}

\newcommand{\rmf}[2]{{\em  Rev. Mex. \ Fiz. \ }{\bf #1}{(#2)}}

\newcommand{\zp}[2]{{\em  Zeitshr. \  Phys. \ }{\bf #1}{(#2)}}
\newcommand{\ib}[2]{{\em  ibid\  }{\bf #1}{(#2)}}

\newcommand{\preprint}[2]{\large
\begin{center}
\hspace*{8cm} Preprint  \, EFUAZ\,  94-{#1}\\
\hspace*{8cm}  {#2} 1994
\end{center}}

\def\@makefnmark{\mbox{$^\@thefnmark$}}

\long\def\@makefntext#1{\noindent$^{\@thefnmark}$#1}
\def\openone{\leavevmode\hbox{\small1\kern-3.3pt\normalsize1}}

\begin{document}
\thispagestyle{empty}

\title{\vspace*{-4cm} \preprint{02}{April} \vspace*{2cm}{\bf  Comment on ''The
Klein-Gordon Oscillator''\linebreak
( S. Bruce and P. Minning, {\bf {\em Nuovo~Cim.}} {\bf 106A}  (1993)
711)}\thanks{Submitted to "Nuovo Cimento A"}}

\author{{\bf Valeri V. Dvoeglazov}\thanks{On leave of absence from {\it Dept.
Theor. and Nucl. Phys., Saratov State University, Astrakhanskaya str., 83,
Saratov RUSSIA}}\,\,\,\thanks{Email: {\it valeri@bufa.reduaz.mx,
dvoeglazov@main1.jinr.dubna.su}}\\
{\it Escuela de F\'{\i}sica, Universidad Aut\'onoma
de Zacatecas}\\
{\it Ciudad Universitaria, Antonio Doval\'{\i} Jaime,\, s/n}\\
{\it  Zacatecas, ZAC., M\'exico\,  98068}}

\date{\empty}
\maketitle

\begin{abstract}
The different  ways of description of the  $S=0$ particle with oscillator-like
interaction are considered.
The results are in conformity with the previous paper of  S.~Bruce and
P.~Minning.
\end{abstract}

\newpage
\setcounter{page}{1}

In connection with the publications of Moshinsky {\it et al.}~\cite{Mosh} the
interest in the
 model with  a  $S=1/2$ Hamiltonian that is linear  in both momenta and
coordinates~\cite{Cook} has grown recently~\cite{oth}.  Analogous type of
interaction
has been considered for the case of $S=0$ and $S=1$ Duffin-Kemmer
field~\cite{DKP} and for the case of $S=0$ Klein-Gordon field~\cite{KG}.

In the paper~\cite{KG} the operators $\vec Q$, coordinate, and $\vec P$,
momentum, have been
represented in $n \otimes n$ matrix form
\begin{equation}
\vec Q=\hat \eta \vec q, \hspace*{1cm} \vec P=\hat \eta \vec p,
\end{equation}
with $\hat \eta^2=1$. The interaction in the Klein-Gordon equation has been
introduced in
the following way:
\begin{equation}
\vec P\rightarrow\vec P-im\hat\gamma\hat\Omega\cdot\vec Q,
\end{equation}
where for the sake of completeness  $\hat \Omega$ is chosen by $3\otimes 3$
matrix
with coefficients $\hat\Omega_{ij}=\omega_{i}\delta_{ij}$ (the  case of
anisotropic oscillator).
The $\hat\gamma$ matrix obeys the following anticommutation relations:
\begin{equation}
\left \{\hat\gamma,\hat\eta\right \}=0, \hspace*{1cm} \hat\gamma^2=1.
\end{equation}
The Klein-Gordon equation for $\Psi (\vec q, t)$, the wave function which could
be expanded in two-component form, is then
\begin{equation}
-\frac{\partial^2}{\partial t^2}\Psi (\vec q, t)= \left (\vec p^{\,2}+m^2\vec
q\cdot\hat\Omega^2\cdot\vec q+m\hat\gamma\,\, tr\Omega+m^2\right )\Psi (\vec q,
t),
\end{equation}
what gives the energy spectrum~\cite{KG}
\begin{eqnarray}
E^2_{(a)N_i}-m^2&=&2m\left (\omega_1 N_1+\omega_2 N_2+\omega_3 N_3\right
),\hspace*{1cm}
N_1, N_2, N_3=0, 1, 2 \ldots\nonumber\\
E^2_{(b)N_i}-m^2&=&2m \left  (\omega_1 (N_1+1)+\omega_2 (N_2+1)+\omega_3
(N_3+1)\right ).
\end{eqnarray}
It  becomes, up to an additive constant,  the  spectrum of the anisotropic
oscillator  in the non-relativistic limit.

However, the physical sense of implementing the matrices $\hat\eta$ and
$\hat\gamma$ in~\cite{KG}  is
obscure. In this short communication  we try to attach some physical
foundations to this procedure.

It is well-known some ways to recast the Klein-Gordon equation in the
Hamiltonian form~\cite{st}-\cite{FV}. First of all, the Klein-Gordon equation
could be re-written to the system
of two coupled equations [8,p.98]
\begin{equation}
\frac{\partial\Psi}{\partial x^{\,\alpha}}=\kappa \Xi_{\,\alpha}, \hspace*{1cm}
\frac{\partial \Xi^{\,\alpha}}{\partial x^{\,\alpha}}=-\kappa\Psi,
\end{equation}
where  $\kappa=mc/\hbar$ (in the following we use the system where
$c=\hbar=1$).
By means of redefining the components
they are easy to present in the matrix Hamiltonian form (cf. with~\cite{KG-d})
\begin{equation}\label{eq:hami}
i\frac{\partial}{\partial t} \left (\matrix{
\phi\cr
\chi_1\cr
\chi_2\cr
\chi_3\cr
}\right )=\left [\left (\matrix{
0 & p_1 & p_2 & p_3 \cr
p_1 & 0 & 0 & 0 \cr
p_2 & 0 & 0 & 0 \cr
p_3 & 0 & 0 & 0 \cr
}\right ) +m\left (\matrix{
1 & 0 & 0 & 0 \cr
0 & -1 & 0 & 0 \cr
0 & 0 & -1 & 0 \cr
0 & 0 & 0 & -1 \cr
}\right )\right ]\left (\matrix{
\phi\cr
\chi_1\cr
\chi_2\cr
\chi_3\cr
}\right ),
\end{equation}
provided that
\begin{eqnarray}
\cases{\phi=i\partial_t \Psi+m\Psi &\cr
\chi_i=-i \bigtriangledown_i \Psi = \vec p_i \Psi .& }
\end{eqnarray}
Using matrices
\begin{equation}
\vec \alpha=\pmatrix{
0_{1\otimes 1} & \mid & {\bf i} & {\bf  j} & {\bf k} \cr
 -  - & -  - & -  - & -  - & -  - \cr
{\bf i}  & \vert & & & \cr
{\bf  j}  & \vert &  &0_{3\otimes 3}& \cr
{\bf k} & \vert &  & &
},\hspace*{1cm} \beta=\pmatrix{
\openone_{1\otimes 1} & \mid & 0 \cr
 - - - & - - - & - - - \cr
&\vert & \cr
0 & \vert & -\openone_{3\otimes 3} \cr
& \vert & \cr
}
\end{equation}
(${\bf \vec i}, {\bf \vec j}, {\bf \vec k}$ are the orth-vectors of  the
Euclidean basis)
and substituting analogously [1a], i.~e., $\vec p\rightarrow \vec
p-im\omega\beta\vec r$,
we come to the equation for upper component
\begin{equation}\label{eq:kgo}
(E^2-m^2)\phi=\left [\vec p^{\,2}+m^2\omega^2\vec r^{\,2}-3m\omega\right ]\phi
,
\end{equation}
what coincides with Eq. (10a) of ref.~\cite{KG} in the case
$\omega_1=\omega_2=\omega_3$.

The similar formulation also originates from the Duffin-Kemmer
approach~\cite{DKP2}.
In this case the wave function  $\Phi=column (\phi_1, \phi_2, \chi_i)$  is
five-dimensional and its components are
\begin{eqnarray}
\cases{\phi_1=(i\partial_t \Psi + m\Psi)/\sqrt{2} & $$\cr
\phi_2=(i\partial_t \Psi - m\Psi)/\sqrt{2} & $$\cr
\chi_i=-i \bigtriangledown_i \Psi = \vec p_i \Psi. & $$}
\end{eqnarray}
It satisfies the equation
\begin{equation}\label{eq:dfk}
i\frac{\partial \Phi}{\partial t}= \left (\vec B \vec p +m\beta_0\right )
\Phi,\hspace*{1cm} B_\mu= [\beta_0, \beta_\mu]_{-}
\end{equation}
(our choice of $5\otimes 5$ dimensional $\beta$-matrices corresponds to ref.
[4b]). As shown there, the substitution $\vec p\rightarrow \vec p - im\omega
\eta_0 \vec r$, where
\begin{eqnarray}
\eta=\pmatrix{
\openone_{2\otimes 2} &\mid & 0 \cr
 -  -  -  & - - -  & - - - \cr
& \vert &\cr
0 & \vert & - \openone_{3\otimes 3}\cr
& \vert &\cr
},
\end{eqnarray}
leads to the equation (\ref{eq:kgo}) for both $\phi_1$ and $\phi_2$. Let us
remark, in both the approaches based on Eq. (\ref{eq:hami}) and the
Duffin-Kemmer  approach, Eq. (\ref{eq:dfk}), we have the equation
\begin{equation}
(E^2 -m^2)\chi_i=(p_i -im\omega x^i) (p_j +im\omega x^j)\chi_j
\end{equation}
for down components,
which seems to  not  be reduced  to oscillator-like equation.

Then, Sakata-Taketani approach~\cite{ST,FV} is characterized by  the equation
which we write
 in the form:
\begin{equation}
i {\partial \over \partial t}\Phi=\left \{\frac{\vec p (\tau_3+i\tau_2) \vec
p}{2m}+m\tau_3\right \} \Phi,
\end{equation}
with $\tau_i$  being the Pauli matrices. $\Phi=column (\phi, \chi)$ is  a
two-component wave function with the components which could be written as
following:
\begin{eqnarray}
\cases{\phi=(\Psi +{i\over m}\partial_t \Psi)/\sqrt{2}& $$\cr
\chi=(\Psi - {i\over m}\partial_t \Psi)/\sqrt{2}.& $$}
\end{eqnarray}

{}From the previous experience we learned that in order to get the
oscillator-like equation we need to do substitution with matrix which
anticommutes with matrix structure of the momentum part of the equation. In our
case the matrix  which has this property  is $\tau_1$ matrix.  Therefore, we do
the substitution
\begin{equation}\label{eq:subs}
\vec p \rightarrow \vec p -im\omega\tau_1 \vec r \nonumber
\end{equation}
and come to
\begin{equation}
E^2 \xi=\left [\vec p^{\,2}+m^2\omega^2\vec r^{\,2}-3m\omega+m^2\right ]\xi,
\end{equation}
where $\xi=\phi+\chi$
and, to the analogous equation for $\eta=\phi-\chi={E\over m}(\phi+\chi)$.  In
the process of calculations we  convinced ourselves that the Hamiltonian
\begin{equation}
{\cal H}={1\over 2m}\left (\vec p^{\,2}+m^2 \omega^2 \vec
x^{\,2}-3m\omega\right ) (\tau_3 +i\tau_2) +m \tau_3
\end{equation}
is the same as in [4b,formula (3.9)] since $\tau_1 (\tau_3 +i \tau_2) \tau_1 =
- (\tau_3 +i\tau_2)$ and $(\tau_3+i\tau_2)\tau_1=\tau_3 + i\tau_2$.

Let us also recall the  Dirac oscillator in $(1+1)$ dimensions~\cite{Doming}.
Curiously, the formula  (\ref{eq:hami}) in $(1+1)$ dimensions  looks like as
the Dirac equation in the case of the choice  of $\gamma$-matrices as
following:
\begin{equation}
\alpha= \tau_x , \hspace*{1cm} \beta=\tau_z.
\end{equation}
In the case of the choice of $\gamma$-matrices as in ref.~\cite{Cooper}, i.~e.,
\begin{equation}
\gamma_0=\tau_1 \hspace*{1cm} \gamma_1=i\tau_3,
\end{equation}
in order to obtain the Dirac oscillator it is necessary to do substitution
(\ref{eq:subs}).

\medskip

I am grateful to A. del Sol Mesa for discussions on the subject of this
communication.


\begin{thebibliography}{99}
\footnotesize{
\bibitem{Mosh} a) M. Moshinsky and A. Szczepaniak, \jp {A22}{17}, L817-L819
(1989)\\
 b) M. Moshinsky, G. Loyola and C. Villegas, in {\em Proc. of the XIII Oaxtepec
Symposium on Nuclear Physics.} Notas de F\'{\i}sica {\bf 13} (1), 187-196
(1990)\\
c) M. Moshinsky et al., in {\em Proc. of  the Rio de Janeiro Int. Workshop on
Relativistic
Aspects of Nuclear Physics.} Eds. T. Kodama et al. World Scientific, Singapore,
1990,
 pp. 271-308\\
d) M. Moshinsky, G. Loyola and A. Szczepaniak, in {\em  Anniversary Volume in
Honour of
J. J. Giambiaggi.} Eds. H. Falomir et al. World Scientific, Singapore, 1990,
pp. 324-349\\
e) C. Quesne and M. Moshinsky, \jp {A23}{12}, 2263-2272 (1990)\\
f) M. Moshinsky, G. Loyola and C. Villegas, \jmp{32}{2}, 373-381 (1991)\\
g) V. I. Kukulin, G. Loyola and M. Moshinsky, \pl {A158}{1-2}, 19-22 (1991)\\
h) M. Moshinsky and G. Loyola, \fp {23}{2}, 197-210 (1993) \\
i)  A. Gonz\'alez, G. Loyola and M. Moshinsky, \rmf {40}{1}, 12-30
(1994)\\[-7mm]

\bibitem{Cook} a) D. It\^o, K. Mori and E. Carrieri, \nc {51A}{4}, 1119-1121
(1967)\\
b) N. V. V. J. Swamy, \pr {180}{5}, 1225-1231 (1969)\\
c) P. A. Cook, \lnc {1}{10}, 419-426 (1971)\\
d) Y. M. Cho, \nc {23A}{3}, 550-556 (1974)\\  [-7mm]

\bibitem{oth} a) M. Moreno and A. Zentella, \jp {A22}{17}, L821-L825 (1989)\\
b) M. Moreno, R. Mart\'{\i}nez and A. Zentella, \mpl {A5}{12}, 949-954 (1990)\\
c) J. Ben\'{\i}tez, R. P. Mart\'{\i}nez y Romero, H. N. N\'u\~nez-Yepez and A.
L. Salas-Brito, \prl {64}{14}, 1643-1645 (1990); \ib {65}{16}, 2085(E)
(1990)\\
d) J. Beckers and N. Debergh, \pr {D42}{4}, 1255-1259 (1990)\\
e) O. Casta\~nos, A. Frank, R. L\'opez and L. F. Urrutia, \pr {D43}{2}, 544-547
(1991)\\
f) R. Mart\'{\i}nez y Romero, M. Moreno and A. Zentella, \pr {D43}{6},
2036-2040 (1991)\\
g) C. Quesne, \ijmp {A6}{9}, 1567-1589 (1991)\\
h) L. M. Jones, Preprint Ill-(TH)-91-24, Univ. of Illinois at Urbana-Champaign,
1991\\
i) V. V. Dixit, T. S. Santhanam and W. D. Thacker, \jmp {33}{3}, 1114-1117
(1992)\\
j) A. Szczepaniak and A. G. Williams, \prd {47}{3}, 1175-1181 (1993)\\
k) J. P. Crawford, \jmp{34}{10}, 4428-4435 (1993)\\
l) V. M. Villalba, Preprint  HEP-TH 9310010, Caracas, Oct. 1993\\   [-7mm]

\bibitem{DKP} a) J. Beckers, N. Debergh and A. G. Nikitin,  \jmp {33}{10},
3387-3392 (1992)\\
b) N. Debergh, J. Ndimubandi and D. Strivay,  \zp {56}{3}, 421-425 (1992)\\
c) Y. Nedjadi and R. C. Barrett, Preprint CNP-93-21, Surrey U., 1993\\   [-7mm]

\bibitem{KG} S. Bruce and P. Minning, \nc {106A}{5}, 711-713 (1993);  \ib
{107A}{1}, 169(E) (1994)\\[-7mm]

\bibitem{st} a) N. Kemmer, \prsl {166}{924}, 127-153 (1938)\\
b) E. C. G. Stueckelberg, {\em Helv. Phys. Acta} {\bf 11}, 225 (1938)\\
[-7mm]

\bibitem{b-st} a) A. March, {\em Quantum Mechanics of Particles and Wave
Fields.} John Wiley. New York, 1951, p.~176\\
b) A. Visconti, {\em Quantum Field Theory. Vol. I.} Pergamon Press,  1969, p.
183\\   [-7mm]

\bibitem{b-st1} E. M. Corson, {\em Introduction to Tensor, Spinors, and
Relativistic
Wave-Equations.} Hafner Pub. Co. New York, 1953\\   [-7mm]

\bibitem{ST} S. Sakata and M. Taketani, {\em Proc. Phys. Math. Soc. (Japan)}
{\bf 22}, 757 (1940)\\   [-7mm]

\bibitem{FV} a) H. Feshbach  and F. Villars, \rmp {30}{1}, 24-45 (1958)\\
b) A. S. Davydov, {\em Quantum Mechanics.} Addison-Wesley Pub. Co., 1968,
section 56, 59\\  [-7mm]

\bibitem{KG-d} M. D. Kostin, \pla {125}{2-3}, 87-88 (1987)\\  [-7mm]

\bibitem{DKP2}a)  R. J. Duffin, \pr{54}{12}, 1114 (1938)\\
b) N. Kemmer, \prsl  {173}{952}, 91-116  (1939)\\  [-7mm]

\bibitem{Doming} F. Dominguez-Adame, \epl {13}{3}, 193-198  (1990); \pl
{A162}{1}, 18-20 (1992)\\[-7mm]

\bibitem{Cooper} F. Cooper {\it et al.}, \apny {187}{1}, 1-28 (1988)
}
\end{thebibliography}
\end{document}